\documentclass[english,showpacs,floatfix]{revtex4}

\usepackage[dvips]{graphicx}
\usepackage{longtable}
\usepackage{amsmath,amssymb}
\usepackage{dcolumn}
\usepackage{latexsym}
\usepackage{babel}
\usepackage[latin1]{inputenc}
\begin{document}
\title{Stability of BTZ black strings}
\vspace{3cm}

\author{Lihui Liu}
\affiliation{Department of Physics, Fudan University, Shanghai
200433, China}

\author{Bin Wang}
\email{wangb@fudan.edu.cn} \affiliation{Department of Physics,
Fudan University, Shanghai 200433, China}


\begin{abstract}
\baselineskip=0.6 cm \centerline{Abstract}

We study the dynamical stability of the BTZ black string against
fermonic and gravitational perturbations. The BTZ black string is
not always stable against these perturbations. There exist
threshold values for $m^2$ related to the compactification of the
extra dimension for fermonic perturbation, scalar part of the
gravitational perturbation and the tensor perturbation,
respectively. Above the threshold values, perturbations are
stable; while below these thresholds, perturbations can be
unstable. We find that this non-trivial stability behavior
qualitatively agrees with that predicted by a thermodynamical
argument, showing that the BTZ black string phase is not the
privileged stable phase.

\end{abstract}

\maketitle

\section{Introduction}

Stability of gravitational configurations has been an intriguing
subject of discussions since long. More recently, the discussion
has been focused on the higher-dimensional black brane solutions
provided by string theories. Gregory and Laflamme have shown in
their pioneering work that neutral black strings in more than four
dimensions suffer from a long-wavelength instability (GL
instability)\cite{1,GLinRS}. In the last few years there have been
a lot of progress achieved in understanding various aspects of the
Gregory-Laflamme instability such as the GL instability with the
presence of charge\cite{GLcharge}, and the GL instability in the
$dS$/$AdS$ spacetimes\cite{GLinRS,Japan}. A review on the black
string instability and whole list of relevant references are
presented in \cite{2}.

Besides dynamical stability, there are also thermodynamical
stability problems of black holes. However for usual black holes,
such as Schwarzschild holes etc, they are thermodynamically
unstable but dynamically stable against perturbations of metric.
The profound connection between dynamical stability and
thermodynamical stability was observed in black branes in
supergravity. Gubser and Mitra proposed a correlated stability
conjecture (GM conjecture) arguing that gravitational systems with
translational symmetry lack stability against small perturbations
often also lack thermodynamical stability\cite{3}. Supports on
this conjecture have been presented in \cite{4,Japan}, while a set
of counterexamples has also been raised \cite{5}. It is of
interest to examine this link further in various configurations.
Recently a non-trivial relation between dynamical and
thermodynamical properties in the background of charged
Kaluza-Klein black hole with squashed horizons was discussed in
\cite{6}.

In this work we are going to study the classical stability of the
BTZ black string solution obtained by starting with a metric
describing an accelerating black hole in $AdS_4$,
\begin{eqnarray}\label{ta1}
    ds^2=\frac{1}{A^2(x-y)^2}\left[P(y)dt^2-\frac{dy^2}{P(y)}
    +\frac{dx^2}{Q(x)}+Q(x)d\varphi^2\right];\\
    \label{ta2}
    P(y)=-\lambda+ky^2-2\bar{m}Ay^3, \ Q(x)=1+kx^2-2\bar{m}Ax^3,\ \ \
\end{eqnarray}
which satisfies the Einstein equation with a negative cosmological
constant
\begin{equation}\label{p2}
    R_{AB}=-(3/l_4^2) g_{AB},
\end{equation}
and $l_4=1/(A\sqrt{\lambda+1})$, $A$ being the acceleration
parameter. The spacetime is sliced onto the branes of
$x/y=\textrm{const.}$, and the visible brane is situated at $x=0$.
There are two configurations of interest:

($i$) The black string phase. To orient with $\bar{m}=0$, the BTZ
black string in $AdS_4$ can be obtained through coordinate
transformation \cite{7}, which is described by
\begin{equation}\label{p1}
    ds^2=l_4^2dz^2+l_4^2 a^2(z) \left[-H^2 dt^2+H^{-2}
    dr^2+r^2d\varphi^2\right],
\end{equation}
where $H^2(r)=\lambda r^2-k$ and $a(z)=\sqrt{\lambda}\cosh z$
being the warp factor. We have the interest on $\lambda>0$ where
locally the constant $z$ slices have the geometry of $AdS_3$. If
$k=+1$, surfaces of constant $z$ have the geometry of BTZ black
hole \cite{8} described by
\begin{equation}\label{p4}
    ds^2_3=-(\lambda r^2-k)dt^2+(\lambda
    r^2-k)^{-1}dr^2+r^2 d\varphi^2,
\end{equation}
which is the solution of $R_{\mu\nu}=-2\lambda g_{\mu\nu}$.

Here, Greek letters take the values $0,1,2$ describing the brane
and the capital Latin letters take the values $0,1,2,3$ describing
the bulk, where $3$ indicates the extra dimension (i.e. $x^0=t,
x^1=r, x^2=\varphi, x^3=z$).

($ii$) The black hole phase. Turn to the case with $m>0$, $Q(x)$
has only one positive root $x_2$. Taking $\Delta
\varphi=4\pi/|Q'(x_2)|$, the conical singularity can be avoided.
The black hole localized on a brane is described by \cite{7}
\begin{equation}\label{ta3}
    ds^2=\frac{1}{A^2}\left[-\left(\lambda r^2-k-\frac{2\bar{m}A}{r}\right)dt^2
    +\left(\lambda
    r^2-k-\frac{2\bar{m}A}{r}\right)^{-1}dr^2+r^2d\varphi^2\right],
\end{equation}
which is the BTZ-like black hole, with extra terms of the form
$2\bar{m}A/r$ coming from the four-dimensional nature of the black
hole. If $2\bar{m}A\ll \lambda^{-1/2}$, the extra term is
negligible outside the horizon, and so the exterior geometry is
essentially identical to that of a BTZ black hole \cite{7}.

A global thermodynamical argument using the entropy was presented
to argue about the stability of these two configurations \cite{7}.
For the BTZ black string, it was shown that its thermodynamic
mass $M_4$ precisely agrees with that of the projected BTZ black
hole $M_3$. The four-dimensional entropy of the black string was
also found in agreement with the three dimensional entropy
\begin{equation}\label{entbs}
S_{BS}=\frac{\pi}{A}\sqrt{\frac{2M_3}{\lambda G_3}},
\end{equation}
where $M_3$ is the projected mass of the BTZ black string on the
visible brane. The black hole entropy was obtained in the form
\begin{equation}\label{entbh}
S_{BH}=\frac{2\pi}{G_4A^2}\frac{\tilde{z}}{\lambda+3\tilde{z}^2+2\tilde{z}^3}.
\end{equation}
The auxiliary variable $\tilde{z}$ is introduced by the identity
$2\bar{m}A=\frac{(\lambda+\tilde{z}^2)\tilde{z}\sqrt{1+\tilde{z}}}{(\lambda-\tilde{z}^3)^{3/2}}$
and noting that $\tilde{z}$ is a monotonic function of $\bar{m}A$,
which leads to $\bar{m}A\rightarrow 0, \tilde{z}\approx
2\bar{m}A\sqrt{\lambda}; \bar{m}A\rightarrow \infty,
\tilde{z}\approx \lambda^{1/3}$.

Equating the black hole and black string entropies,
Eq.(\ref{entbs}) and Eq.(\ref{entbh}), and identifying the mass of
the two phases by substituting
$M_3=\frac{1}{2G_3}\frac{\tilde{z}^2(1+\tilde{z})(\lambda-\tilde{z}^3)}
{(\lambda+3\tilde{z}^2+2\tilde{z}^3)^2}$, we
get a fourth order polynomial in terms of the auxiliary variable
$\tilde{z}(\lambda-\tilde{z}^2-\tilde{z}^3)=0$. The root
$\tilde{z}=0$ simply indicates that the mass and entropy of the
localized BTZ black hole (when $\bar{m}A\rightarrow 0$) match
that of the black string, which shows that thermodynamically there
is no privileged stable phase in these two configurations. When
$\bar{m}A$ increases so that the auxiliary variable $\tilde{z}$
reaches $\tilde{z}_x$, which satisfies
$\tilde{z}_x^2+\tilde{z}_x^3=\lambda$, the black hole and the
black string entropies equal to each other again. Except  these
two points, the entropy comparison of black hole and black
string phase is not illuminating in the auxiliary variable
$\tilde{z}$. However expressing the entropies of the black hole
and the black string phase in terms of their masses, it was shown
in Fig.1 of \cite{7} that, for small $\bar{m}A$, the black string
phase entropy exceeds that of the localized black hole,
while for large $\bar{m}A$, the result goes in the contrary; the
changeover point is at the root $\tilde{z}_x$.

In the present work, we will study the stability of the BTZ black string
against fermionic and gravitational perturbations and examine the
relation between the dynamical stability against different fields'
perturbations and the thermodynamical stability observed in
\cite{7}. Since we have, in this background, exact solutions for
both the black strings in 3+1 dimensions as well as BTZ black
holes on the brane in 2+1-dimensions, we can examine the stability
of the black string in detail.

\section{Fermion Perturbation}

In this section we take a first look at the stability of the BTZ
black string against fermionic perturbation. We assume that the
Fermion field exists in the bulk. We will derive the fermionic
perturbation equation and examine the mode which signals the
stability of the black string.

\subsection{Perturbation Equation}

For the massless Fermion field, it is described by the Dirac
equation\cite{Deq}
\begin{equation}\label{fe1}
    \gamma^ae^{A}_a(\partial_{A}+\Gamma_{A})\Psi=0.
\end{equation}
Here $\gamma^a$ is the Dirac matrix satisfying
$\{\gamma^a,\gamma^b\}=2\eta^{ab}$, $\eta^{ab}$ is the Lorentz
metric with the signature $(-1,1,1,1)$. $e^A_a$ is the tetrad so
chosen that $e^A_a e^B_b \eta^{ab}=g^{AB}$, whose indices in small
letters are raised or lowered by $\eta_{ab}$ and indices in
capital letters are raised or lowered by $g_{AB}$.  $\Gamma_A$ is
the affinity of covariant derivative, defined by
$\Gamma_A=\frac{1}{8}[\gamma^a,\gamma^b]e^B_a e_{bB;A}$,
$e_{bB;A}=e_{bB,A}-\Gamma^{E}_{BA}e_{bE}$. In the spacetime
background of Eq.(\ref{p1}), we can rewrite Eq.(\ref{fe1}) as
\begin{equation}\label{fe2}
    \frac{\gamma^0}{H}  \partial_t \Psi +\gamma^1
    H\left[\partial_r +\frac{(2\lambda r^2-k)}{2r(\lambda
    r^2-k)}\right]\Psi+\frac{\gamma^2}{r}\partial_{\varphi}\Psi
    +\gamma^3a\left(\partial_z+\frac{3\partial_za}{2a}\right)\Psi=0.
\end{equation}
Separating the wave function by
\begin{equation}\label{fe00}
    \Psi(t,r,\varphi,z)=R(r)Z(z)e^{-i\omega
    t+i\mu\varphi},
\end{equation}
and taking
\begin{eqnarray}\label{fe01}
    &&R(r)=[r^2(\lambda r^2-k)]^{-1/4}\Phi(r),\\
\label{fe02}
    &&Z(z)=\exp\left[-\int{\frac{3\partial_z
    a}{2a}dz}+im\int{\frac{dz}{a}}\right],
\end{eqnarray}
we can obtain the radial wave equation in the form
\begin{equation}\label{fe3}
    -i\omega H^{-1}\gamma^0 \Phi+H\gamma^1 \partial_r\Phi+i(\mu r^{-1}
    \gamma^2+m\gamma^3)\Phi=0.
\end{equation}

Here, $\mu$ takes the value of $(\textrm{integer times}\
2\pi)/(\textrm{period of}\ \varphi)$ as required by the periodical
condition in $\varphi$. The value of $m$ is determined by the
compactification of the extra-dimension.

Choosing
\begin{equation}\label{fe4}
    \gamma^0=\left(\begin{array}{cc} 0 & -i \\ -i & 0 \\
    \end{array}\right),\ \
    \gamma^i=\left(\begin{array}{cc} 0 & -i\sigma^i \\ i\sigma^i & 0 \\
    \end{array}\right),\ \ i=1,2,3,
\end{equation}
and letting $\Phi=(\phi_1,\phi_2)^\textrm{T}$,
$\phi_a=(\phi_{a+},\phi_{a-})^\textrm{T}$ $(a=1,2)$, we have
\begin{eqnarray}
\label{fe5}
  i\omega H^{-1}\phi_1+H\sigma^1
  \partial_r\phi_1+i(\mu r^{-1}
    \sigma^2+m\sigma^3)\phi_1&=&0,\\
\label{fe6}
  -i\omega H^{-1}\phi_2+
  H\sigma^1 \partial_r\phi_2+i(\mu r^{-1}
    \sigma^2+m\sigma^3)\phi_2&=&0.
\end{eqnarray}
We focus on the second equation and the first one can be solved
easily by changing $\omega \rightarrow -\omega$. Explicitly
writing out the Pauli matrices, Eq.(\ref{fe6}) gives
\begin{eqnarray}\label{fe7}
\left\{%
\begin{array}{c}
    (-i\omega H^{-1}+im)\phi_{2+}+(H\partial_r
  -\mu r^{-1})\phi_{2-} = 0, \\
    (-i\omega H^{-1}-im)\phi_{2-}+(H\partial_r
  +\mu r^{-1})\phi_{2+} = 0.
\end{array}%
\right.
\end{eqnarray}
Defining $Y_{\pm}=\left(\frac{rm+i\mu}{rm-i\mu}\right)^{\pm
1/4}(\phi_{2+}\pm\phi_{2-})$, we can change these equations into
\begin{equation}\label{fe9}
  (\partial_{\bar{r}}\mp S)Y_{\pm}\pm im Y_{\mp}=0,
\end{equation}
where
\begin{eqnarray}\label{fe10}
    &&\ \ \bar{r}=\int{\frac{\sqrt{r^2 m^2+\mu^2}}{m r
    H}dr},\\
    \label{fe11}
    &&S(r)=\frac{im
    r}{H\sqrt{r^2m^2+\mu^2}}\left[\omega
    -\frac{\mu m H^2}{2(r^2m^2+\mu^2)}\right].
\end{eqnarray}
Decoupling Eqs.(\ref{fe9}), we have
\begin{equation}\label{fe12}
    -\partial^2_{\bar{r}}Y_{\pm}+(m^2+S^2\pm\partial_{\bar{r}}S)Y_{\pm}=0.
\end{equation}
Eq.(\ref{fe9}) and Eq.(\ref{fe12}) allow us to look at the
asymptotic behavior of $Y_{\pm}$ at the horizon
$r=r_H=\sqrt{k/\lambda}$ and at spatial infinity $r\rightarrow
\infty$. In the vicinity of the horizon $r\rightarrow r_H+0$,
$Y_{\pm}\sim (r-r_H)^{\rho_{\pm}}$, and from Eq.(\ref{fe12}) we
require $\rho_{\pm}$ to satisfy
$2\rho_{\pm}^2-\rho_{\pm}+\omega^2/2\lambda k\pm
i\omega/2\sqrt{\lambda k}=0$. Since there exist only ingoing
waves near the horizon, we have $\rho_+=1/2-i\omega/2\sqrt{\lambda
k}$, and $\rho_-=-i\omega/2\sqrt{\lambda k}$ and thus in the limit
$r\rightarrow r_H+0$,
\begin{equation}\label{fe14}
    Y_+\sim(r-r_H)^{\frac{1}{2}-\frac{i\omega}{2\sqrt{\lambda k}}},\ \
    Y_-\sim(r-r_H)^{-\frac{i\omega}{2\sqrt{\lambda
    k}}}.
\end{equation}
For a consistency check this result also satisfies the coupled
first-order equation (\ref{fe9}).  Similarly we let $r\rightarrow
\infty$ and find asymptotic solutions satisfying Eq.(\ref{fe9})
and Eq.(\ref{fe12})
\begin{equation}\label{fe15}
    Y_{\pm}\sim \left\{
\begin{array}{c}
  r^{m/\sqrt{\lambda}} \\
  r^{-m/\sqrt{\lambda}} \\
\end{array}
\right\},\ \  r\rightarrow \infty.
\end{equation}

\subsection{Quasinormal Frequencies of Fermionic Perturbation}

We now investigate the stability problem by solving
Eq.(\ref{fe9}), which is similar to the quasinormal modes (QNMs)
study (a review on the QNMs study can be found in \cite{9,10}).

In $AdS$ spacetime, the QNMs were defined in \cite{HoroQNM} by
only ingoing modes near the horizon while at the spatial infinity
the wave function $\Psi$, here equivalent to $r^{-1}Y_{\pm}$,
vanishes. This boundary condition allows only a discrete set of
complex $\omega$ to exist. If the imaginary part of the QNM
frequency is negative, we will have the decaying mode. However, if
the imaginary part of QNM frequency is positive, we will have a
growing mode, indicating that the spacetime is unstable against
the field perturbation.

To calculate the QNM frequencies, we need to solve Eq.(\ref{fe9})
and select only solutions which satisfy $\Psi\rightarrow 0$ as
$r\rightarrow \infty$. For simplicity, we will let $\mu=0$, which
allows to solve Eq.(\ref{fe9}) analytically. Working with the
decoupled equation Eq.(\ref{fe12}), we can rewrite them as
\begin{equation}\label{drs1}
    H^2\partial^2_rY_-+\lambda r\partial_rY_-+\left(m^2-H^2+i\omega H^{-2}\lambda
    r\right)Y_-=0.
\end{equation}
This equation is symmetric under the transformation of
$r\rightarrow \alpha r$, $m\rightarrow m/\alpha$ and
$\omega\rightarrow \omega/\alpha$. In the following we will set
$\lambda=1$. Taking $x=(r-1)/(r+1)$,
$Y_-=x^{-i\omega/2}(x-1)^{-m}y(x)$, Eq.(\ref{drs1}) becomes a
hypergeometric equation (see also \cite{alswq})
\begin{equation}\label{drs2}
    x(1-x)\frac{d^2y}{dx^2}+[c-(1+a+b)x]\frac{dy}{dx}-ab\ y=0,
\end{equation}
where $a=-m$, $b=1/2-i\omega-m$, and $c=1/2-i\omega$. We choose
its solution to be simply $_2F_1(a,b,c;x)$, and eliminate the
other one $x^{1-c}\ _2F_1(a-c+1,b-c+1,2-c;x)$, so that $Y_-$
contains only ingoing modes near the horizon embodied by the
factor $x^{-i\omega/2}$. Therefore $Y_-=x^{-i\omega}(x-1)^{-m}\
_2F_1(a,b,c;x)$. Considering $_2F_1(a,b,c;x)=(1-x)^{c-a-b}\
_2F_1(c-a,c-b,c;x)$ \cite{Int}, we can rewrite the solution as
\begin{equation}\label{drs3}
Y_-=x^{-i\omega/2}(x-1)^{-|m|}\ _2F_1(\tilde{a},\tilde{b},c;x),
\end{equation}
where $\tilde{a}=-|m|$ and $\tilde{b}=1/2-i\omega-|m|$ satisfying
$\textrm{Re}(c-\tilde{a}-\tilde{b})=2|m|>0$. We need to substitute
Eq.(\ref{drs3}) into the first equation of Eqs.(\ref{fe9}) to get
the explicit expression of $Y_+$. Using $\frac{d}{dx}\
_2F_1(a,b,c;x)=\frac{ab}{c}\ _2F_1(a+1,b+1,c+1;x)$ \cite{Int}, we
have
\begin{equation}\label{drs4}
    Y_+=-i\epsilon\ x^{1/2-i\omega/2}(x-1)^{-|m|}\left[\
      _2F_1(\tilde{a},\tilde{b},c;x)
    +C\ (x-1)\ _2F_1(\tilde{a}+1,\tilde{b}+1,c+1;x)\right],
\end{equation}
where $\epsilon$ is the sign of $m$ and
$C=\frac{1-2i\omega-|m|}{1-2i\omega}$. Using the transformation
\begin{eqnarray}\nonumber
    &_2F_1(a,b,c;x)=\frac{\Gamma(c)\Gamma(c-a-b)}{\Gamma(c-a)\Gamma(c-b)}\
    _2F_1(a,b,a+b-c+1;1-x)\\ &\ \ \ \ \
    \ \ \ \ \ \ \ \ \
    \ \ \ +(1-x)^{c-a-b}\frac{\Gamma(c)\Gamma(a+b-c)}{\Gamma(a)\Gamma(b)}\
    _2F_1(c-a,c-b,c-a-b+1,1-x),\label{drs5}
\end{eqnarray}
and the $x=\frac{r-1}{r+1}$ or $x-1\sim -2/r$ at spatial infinity,
we can write the asymptotical form of $Y_{\pm}$ at spatial
infinity as
\begin{eqnarray}\label{drs006}
    &&Y_+\sim i\epsilon \left(D_+r^{|m|}+D_-r^{-|m|}\right),\\
    &&Y_-\sim D_+r^{|m|}+D_-r^{-|m|},
\end{eqnarray}
where
\begin{equation}\label{drs007}
    D_+=(-2)^{-|m|}\frac{\Gamma(c)\Gamma(c-\tilde{a}-\tilde{b})}{\Gamma(c-\tilde{a})\Gamma(c-\tilde{b})},\
    \ \
    D_-=(-2)^{|m|}\frac{\Gamma(c)\Gamma(\tilde{a}+\tilde{b}-c)}{\Gamma(\tilde{a})\Gamma(\tilde{b})}.
\end{equation}
This matches the asymptotical behavior already obtained in
Eq.(\ref{fe15}).

When $|m|<\sqrt{\lambda}$, it is easy to see that the boundary
condition $r^{-1}Y_{\pm}\rightarrow 0$ at spatial infinity can be
satisfied automatically, thus we have no limit from the boundary
condition requirement on the value of $\omega$. This is quite
unusual since the QNM with any frequency, even with the positive
imaginary part can be allowed. This implies that when
$|m|<\sqrt{\lambda}$, the perturbation may have growing modes and
the black string is unstable against the Fermonic perturbation.

On the other hand for the perturbations of
$|m|\geq\sqrt{\lambda}$, the fulfillment of boundary condition
$r^{-1}Y_{\pm}\rightarrow 0$ at spatial infinity calls for the
condition that $D_+=0$. Thus we need to let $c-\tilde{a}$ be zero
or negative integer, which gives the discrete values of QNM
frequencies
\begin{equation}\label{drs6}
    \omega=-i\left(1/2+n+|m|/\sqrt{\lambda}\right), \ \ n=0,1,2,...
\end{equation}
where we have recovered $\lambda$. The purely negative imaginary
frequencies indicate that when $|m|\geq\sqrt{\lambda}$ the
perturbation of massless Dirac field exhibits the decay mode
showing that the black string is stable under such perturbation.

\section{Gravitational Perturbation}

Now we consider the linear perturbation of the BTZ black string
spacetime, which we will denote as $g_{AB}+h_{AB}$ where $g_{AB}$
stands for the components of the unperturbed black string metric
and $h_{AB}$ is the metric perturbation.

\subsection{Perturbation equations}

The perturbation equations can be greatly simplified by choosing
appropriate gauge conditions, which was discussed in detail in
studying the gravitational wave in asymptotically flat spacetimes
(e.g., \cite{Wd}, \S 4.4). We will employ gauge conditions on
$h_{AB}$ including the tracelessness condition, the transversality
condition, namely $h^A_A=0$, $h^{AB}_{;B}=0$, and $h_{4\mu}=0$, so
that
\begin{equation}\label{p3}
    h_{AB}(t,r,\varphi,z)=e^{\Omega t-i\mu\varphi}\left(
             \begin{array}{cccc}
               H_{tt}(r,z) & H_{tr}(r,z) & H_{t\varphi}(r,z) & 0 \\
               H_{rt}(r,z) & H_{rr}(r,z) & H_{r\varphi}(r,z) & 0 \\
               H_{\varphi t}(r,z) & H_{\varphi r}(r,z) &
               H_{\varphi\varphi}(r,z) & 0 \\
               0 & 0 & 0 & H_{zz}(r,z) \\
             \end{array}
           \right).
\end{equation}
The linearized Ricci tensor has the form $\delta
R_{BD}=-\frac{1}{2}\square^{(4)}
h_{BD}+R_{E(B}h^E_{D)}-R_{BCDE}h^{CE}$. Applying this to linearize
Eq.(\ref{p2}), we get
\begin{equation}\label{p8}
    \square^{(4)}h_{BD}+2R_{BCDE}h^{CE}=0.
\end{equation}
Inserting the perturbation Eq.(\ref{p3}) into the above equation,
we find immediately that with the trace-free condition, the
equation governing $H_{zz}$ is decoupled from the rest, giving the
description of scalar part of the perturbation. Taking the ansatz
$H_{zz}(r,z)=H_{zz}(r)u_m(z)$, we have
\begin{equation}\label{p5}
    (\lambda r^2-k)H_{zz,rr}+r^{-1}(3\lambda
    r^2-k)H_{zz,r}-[(\lambda
    r^2-k)^{-1}\Omega^2-2\lambda+\mu^2 r^{-2}+m^2]H_{zz}=0
\end{equation}
where $m^2$ is the effective mass of the Kaluza-Klein modes
satisfying
\begin{equation}\label{p7}
    a^{-1}(a^3u_m')'-6(a')^2u_m+m^2u_m=0,
\end{equation}
where the prime denotes the derivative with respect to $z$. In the
vicinity of the horizon and near the spatial infinity, solutions
of Eq.(\ref{p5}) asymptotically have behaviors near the horizon
and the spatial infinity
\begin{eqnarray}\label{p15}
    &&H_{zz}\sim (r-r_H)^{\pm\frac{\Omega}{2\sqrt{\lambda k}}},\ \
    r\rightarrow r_H, \\ \label{p16}
    &&H_{zz}\sim r^{-1\pm\sqrt{m^2/\lambda-1}},\ \ r\rightarrow \infty.
\end{eqnarray}
$r_H=\sqrt{k/\lambda}$ denotes the event horizon. In order to have
only ingoing wave near the horizon, we require the sign in the
exponential of Eq.(\ref{p15}) to be positive.

Now we turn to discuss the tensor perturbation. Since $H_{zz}$ is
independent of the rest, i.e., of the tensor perturbation, we set
$H_{zz}\equiv0$. However to obtain the perturbation equation, we
need to take further simplification. Considering just the s-wave
\cite{1} that $H_{t\varphi}=0$ and $H_{r\varphi}=0$, we have
\begin{equation}\label{p6}
    h_{AB}(t,r,z)=e^{\Omega t}\left(
             \begin{array}{cccc}
               H_{tt}(r,z) & H_{tr}(r,z) & 0 & 0 \\
               H_{rt}(r,z) & H_{rr}(r,z) & 0 & 0 \\
               0 & 0 & H_{\varphi\varphi}(r,z) & 0 \\
               0 & 0 & 0 & 0 \\
             \end{array}
           \right).
\end{equation}

Using the metric Eqs.(\ref{p1})(\ref{p4}) and the perturbation in
the form (\ref{p6}), we find that the perturbation equation
reduces to
\begin{equation}\label{p9}
    a^{-2}\left(\square^{(3)}h_{\mu\nu}+2R^{(3)}_{\mu\lambda\nu\rho}
      h^{\lambda\rho}\right)
    +a\left(a^{-1}h'_{\mu\nu}\right)'-2a^{-1}a''h_{\mu\nu}=0,
\end{equation}
where the prime denotes the derivative with respect to $z$, and
the greek index relates to the three-dimensional spacetime
Eq.(\ref{p4}). Separating the perturbation into
$h_{\mu\nu}(t,r,z)=h_{\mu\nu}(t,r)v_m(z)=e^{\Omega
t}H_{\mu\nu}(r)v_m(z)$, we can rewrite (\ref{p9}) into
\begin{equation}\label{p10}
    a\left(a^{-1}v_m'\right)'-2a^{-1}a''v_m+a^{-2}m^2v_m=0,
\end{equation}
and
\begin{equation}\label{p11}
    \square^{(3)}h_{\mu\nu}+2R^{(3)}_{\mu\lambda\nu\rho}
    h^{\lambda\rho}-m^2h_{\mu\nu}=0.
\end{equation}
With the aid of gauge conditions, we can obtain a second order
differential equation of $H_{tr}$ from Eq.(\ref{p11}) through
tedious calculations, which reads
\begin{eqnarray}\label{p12}
    &&\ \ \ \ \left[\Omega^2+(\lambda r^2-k)m^2+\lambda^2 r^2-2\lambda
    k\right]H_{tr,rr} \nonumber\\
    &&+\left[\frac{7\lambda r^2-k}{r(\lambda r^2-k)}
      \Omega^2+\frac{5\lambda r^2-k}{r}m^2
    +\frac{\lambda(5\lambda^2 r^4-13\lambda k r^2+2k^2)}{r(\lambda
    r^2-k)}\right]H_{tr,r} \nonumber \\
    &&+\left[\frac{\lambda(3\lambda r^2-2k)(\lambda^2r^4-4\lambda k r^2-k^2)}
    {r^2(\lambda r^2-k)^2}+\frac{2\lambda^2 r^4-2\lambda  k
      r^2-k^2}{r^2(\lambda r^2-k)}m^2
    \nonumber \right. \nonumber \\
    &&\left.+\frac{6\lambda^2 r^4-k^2}{r^2(\lambda r^2-k)^2}\Omega^2
    -\left(m^2+\frac{\Omega^2}{\lambda
    r^2-k}\right)^2\right]H_{tr}=0.
\end{eqnarray}
Its asymptotic behaviors near the horizon and the spatial infinity
have forms as
\begin{eqnarray}\label{p13}
    &&H_{tr}\sim (r-r_H)^{-1\pm\frac{\Omega}{2\sqrt{\lambda k}}},\ \
    r\rightarrow r_H, \\ \label{p14}
    &&H_{tr}\sim r^{-2\pm\sqrt{m^2/\lambda+1}},\ \ r\rightarrow \infty.
\end{eqnarray}
Considering only the ingoing wave near the horizon, we will just
keep the positive sign in the exponent of Eq.(\ref{p13}).

We have obtained equations describing the scalar part of the
gravitational perturbation (\ref{p5}) and the tensor perturbation
(\ref{p12}). In the following we will adopt the boundary condition
that at the spatial infinity the wave functions $H_{zz}$ and
$H_{tr}$ vanish so as to solve these equations and examine the
stability of the BTZ black string against gravitational
perturbations.

\subsection{Exact Solution of Scalar Part Perturbation}

Taking $H_{zz}(r)=r^{-1/2}\psi(r)$ and $\omega=i\Omega$,
Eq.(\ref{p5}) can be put in the form of a Schr\"{o}dinger
equation,
\begin{equation}\label{e1}
    \frac{d^2\psi}{dr^2_*}+[\omega^2-V(r)]\psi=0,
\end{equation}
where $r_*$ is the tortoise coordinate defined by $r_*=\int
dr/(\lambda r^2-k)=\frac{1}{2\sqrt{\lambda k}} \ln\frac{r-r_H}
{r+r_H}$, such that the horizon is at $r_*=-\infty$ and the
spacial infinity locates at $r_*=0$. The effective potential reads
\begin{equation}\label{e2}
    V(r)=(\lambda r^2-k)[m^2-5\lambda/4+(k/4+\mu^2)r^{-2}].
\end{equation}
For simplicity we will take $\lambda=1$. The potential is very
similar to that of the massless scalar field perturbation
described by the Klein-Gorden equation in the BTZ black hole on
the brane where $V(r)=(\lambda
r^2-k)[m^2-5\lambda/4+(k/4+\mu^2)r^{-2}]$\cite{BTZQNM1}, which
vanishes at the horizon but diverges at spatial infinity.

Adopting $x=1/\cosh^2 r_*=(r^2-r^2_H)/r^2$\cite{BTZQNM1,BTZQNM},
and $\psi(r(x))=(1-x)^{1/4-\sqrt{m^2-1}/2}x^{-i\omega/2}y(x)$, we
can transform Eq.(\ref{e1}) into a hypergeometric equation as
Eq.(\ref{drs2}),  but now with $a =
\frac{1}{2}(1-\sqrt{m^2-1}-i\omega+i\mu)$, $b =
\frac{1}{2}(1-\sqrt{m^2-1}-i\omega-i\mu)$, and $c = 1-i\omega$. To
meet the boundary condition at the horizon $x=0$ where there are
only ingoing waves, we choose the solution of the hypergeometric
equation to be simply $y(x)=_2F_1(a,b,c;x)$, so that
\begin{eqnarray}
\label{e4}
    \psi=(1-x)^{1/4-\sqrt{m^2-1}/2}x^{-i\omega/2}\
    _2F_1(a,b,c;x),\\
\label{e41} H_{zz}=(1-x)^{1/2-\sqrt{m^2-1}/2}x^{-i\omega/2}\
    _2F_1(a,b,c;x).
\end{eqnarray}
We will check the restrictions of the boundary condition at
spatial infinity. We write the asymptotical form of $H_{zz}$ at
spatial infinity in the standard form Eq.(\ref{p16}). Using the
transformation relation of hypergeometric function Eq.(\ref{drs5})
we have, at spatial infinity,
\begin{equation}\label{e0111}
    H_{zz}\sim
    G_+r^{-1+\sqrt{m^2-1}}+G_-
    r^{-1-\sqrt{m^2-1}},
\end{equation}
where
\begin{equation}\label{e0112}
    G_+=\frac{\Gamma(c)\Gamma(c-a-b)}{\Gamma(c-a)\Gamma(c-b)},\ \
    \ G_-=\frac{\Gamma(c)\Gamma(a+b-c)}{\Gamma(a)\Gamma(b)}.
\end{equation}

When $m^2<2\lambda$, it is easy to see that the boundary condition
that $H_{zz}$ vanish at infinity is satisfied automatically, which
is similar to what we observed in the fermonic perturbation. This
means that there is no limit on the frequency of the perturbation,
which can allow the perturbation with growing modes to exist and
make the spacetime to be unstable.

When $m^2\geq 2\lambda$, we need $G_+=0$ to ensure the vanishing
of $H_{zz}$ at spatial infinity. Thus we should have $c-a$ or
$c-b$ to be zero or negative integers, which leads to
\begin{equation}\label{e5}
    \omega=\pm \mu-i(2n+1+\sqrt{m^2/\lambda-1}),
\end{equation}
where we have restored the $\lambda$. Notice that the imaginary
part of the frequency is negative, which describes the decay mode
of the perturbation. Thus the BTZ black string is stable in this
case.

\subsection{Numerical Result of the Tensor Perturbation}

Now we try to solve the equation of the tensor perturbation
Eq(\ref{p12}). It has to satisfy the boundary condition that
$H_{tr}$ vanishes at spatial infinity.

From the asymptotic behavior Eq.(\ref{p14}), it is easy to see
that the boundary condition can be satisfied automatically when
$m^2<3\lambda$. This again means that in this situation, there is
no restriction on the frequency of the perturbation and even the
growing mode is permitted which can make the spacetime unstable.

When $m^2\geq 3\lambda$, we need to solve Eq.(\ref{p12}) to
examine the perturbation behavior. This equation cannot be solved
analytically, which requires numerical calculation. Considering
that the spacetime has the AdS property, we will employ the
numerical technique developed in \cite{HoroQNM} and applied in
many AdS black hole spacetimes \cite{bin}. The radial equation
Eq(\ref{p12}) keeps the same form under the transformation
$r\rightarrow \alpha r$, $m\rightarrow m/\alpha$ and
$\Omega\rightarrow \Omega/\alpha$. We will take $\lambda=1$ for
simplicity. Taking $x=r^{-1}$, we can rewrite Eq.(\ref{p12}) in
the form
\begin{equation}\label{n1}
    s(x)H_{tr,xx}+\frac{t(x)}{x-x_H}H_{tr,x}+\frac{u(x)}{(x-x_H)^2}H_{tr}=0,
\end{equation}
where $x_H=1/r_H=1$, and
\begin{eqnarray}
  s(x) &=& (2+m^2-\Omega^2)x^6+(4+2m^2-2\Omega^2)x^5-2(1+m^2)x^3-(1+m^2)x^2, \label{n2}\\
  t(x) &=& (2+m^2-\Omega^2)(x^6+x^5)-(7+2m^2-5\Omega^2)(x^4+x^3)-3(1+m^2)(x^2+x), \label{n3}\\
  u(x) &=& -(2+m^2-\Omega^2)x^6-(5 + m^2 - m^4 + 2 m^2 \Omega^2 - \Omega^4)x^4\nonumber\\
  &&+2(7+2m^2-m^4-3\Omega^2+m^2\Omega^2)x^2-3-2m^2+m^4.  \label{n4}
\end{eqnarray}
We can expand these functions around $x_H$, e.g.
$s(x)=\sum_{n=0}^{\infty}s_n(x-x_H)^n$, and so does $t(x)$ and
$u(x)$. $H_{tr}$ can also be written in the expansion form
$H_{tr}=\sum_{n=0}^{\infty}a_n(x-x_H)^{n+\rho}$. Inserting all
these expansions into Eq.(\ref{n1}), we can solve the leading
order equation and find $\rho=-1\pm \Omega/2$. We will take the
plus sign which stands for the in-going wave near the horizon.
Equating the coefficients of $(x-x_H)^n$, we have the recursion
relation for $a_n$:
\begin{equation}\label{n5}
    a_n=-\frac{1}{P_n}\sum_{k=0}^{n-1}[(k+\rho)(k+\rho-1)s_{n-k}
    +(k+\rho)t_{n-k}+u_{n-k}]a_k,
\end{equation}
where
\begin{equation}\label{n6}
    P_n=(n+\rho)(n+\rho-1)s_0+(n+\rho)t_0+u_0.
\end{equation}
Considering the boundary condition that $H_{tr}|_{x=0}=0$ at
infinity, we require that
$\sum_{n=0}^{\infty}a_n(\Omega,m)(-x_H)^n=0$, which is satisfied
only for discrete values of $\Omega$. In the later discussion, we
will take $\omega=i\Omega$. We need to find the zeros of
$\sum_{n=0}^{\infty}a_n(\omega,m)(-x_H)^n=0$ in the complex
$\omega$ plane. This is done by truncating the series after a
large number of terms and computing the partial sum as a function
of $\omega$. We can then find zeros of this partial sum, and check
the accuracy by seeing how much the location of the zero changes
as we go to higher partial sums.

To check the numerical scheme, we will first use it to calculate
the scalar part perturbation, which can be compared with
analytical results. For the scalar part, the corresponding
$s(x)=4x^4+8x^3+4x^2$, $t(x)=8x^4+8x^3$, and
$u(x)=(1+4\mu^2)x^4-(6-4m^2+4\mu^2+4\Omega^2)x^2+5-4m^2$. We
truncate the infinite series at $N=200$ and get the numerical
result listed in Table 1. We find that the numerical calculation
agrees well with the analytic solution, especially for the lower
modes. This gives us confidence to use the numerical scheme to
explore the tensor perturbation. Results for the tensor
perturbation are shown in Table 2. It is interesting to find that
for $m^2\geq 3\lambda$, all modes have negative imaginary parts of
QNM frequencies, which shows that in this case the spacetime is
stable against tensor perturbation. Furthermore we observed that
for lower modes the QNM frequency only has the imaginary part in
the tensor perturbation and surprisingly its value satisfies
$\omega=-i(2n-1+\sqrt{m^2+1})$, $n=0,1,2...$.

\begin{center}
\begin{tabular}{ccc|rr}
  \hline\hline
  {\ \ $m^2$\ \ } & {\ \ $\mu$\ \ } & {\ \ $n$\ \ } &
  {\ \ \ \ \ \ \ \ \ \ \ \ $\omega_{\textrm{num}}$\ \ \ \ \ } &
  {\ \ \ \ \ \ \ \ \ \ \ \ $\omega_{\textrm{exact}}$\ \ \ \ \ } \\
  \hline
  $2$ & 0 & 0 & $-2.0000i$ & {$-2i$\ \ } \\
    &   & 1 & $-4.0001i$ & {$-4i$\ \ } \\
    &   & 2 & $-6.0002i$ & {$-6i$\ \ } \\
    &   & 3 & $-8.0003i$ & {$-8i$\ \ } \\
    \hline
   $2$ & 1 & 0 & $\pm1.0000-2.0000i$ & {$\pm1-2i$\ \ } \\
    &   & 1 & $\pm1.0000-4.0000i$ & {$\pm1-4i$\ \ } \\
    &   & 2 & $\pm0.97-5.997i$ & {$\pm1-6i$\ \ } \\
   \hline
   $3$ & 0  & 0 & $-2.4140i$ & {$-2.41421i$\ \ } \\
     &    & 1 & $-4.4142i$ & {$-4.41421i$\ \ } \\
     &    & 2 & $-6.41i$ & {$-6.41421i$\ \ } \\
  \hline
\end{tabular}
\end{center}
\vskip 1mm \begin{quote}\small{ Table 1. Comparison between the
numerical result and the analytic result of $\omega$ for the
scalar part perturbation. }
\end{quote}

\begin{center}
\begin{tabular}{r|rrr|r}
  \hline\hline
  {\ \ \ \ \ $m^2$\ \ } & {\ \ \ \ \ \ \ \ \ \ \ \ $\omega|_{n=0}$}
  & {\ \ \ \ \ \ \ \ \ \ \ \ $\omega|_{n=1}$}
  & {\ \ \ \ \ \ \ \ \ \ \ \ \ \ \ \ \ \ $\omega|_{n=2}$\ \ \ \ \ \ \ \ \ } &
   {\ \ $-1+\sqrt{m^2+1}$\ \ }\\
  \hline

  $3.001\ \ $ & $-1.0002500i$ & $-3.0002500i$ & $\pm
  2\times10^{-7}-5.0002500i\ \ $ & $1.0002500\ \ $ \\
  $4\ \ $ & $-1.236068i$ & $-3.236068i$ & $\pm 0.00007-5.236068i\ \ $ &
  $1.2360680\ \ $ \\
  $5.25\ \ $ & $-1.5000000i$ & $-3.5000000i$ & $\pm 0.00004-5.5000000i\ \
  $ & $1.5\ \ $ \\
  $6.25\ \ $ & $-1.6925824i$ & $-3.6925824i$ & $\pm 0.00001-5.6925824i\ \
  $ & $1.6925824\ \ $ \\
  $8.01\ \ $ & $-2.0016662i$ & $-4.0016662i$ &
  $\pm3\times10^{-8}-6.0016662i\ \ $ & $2.0016662\ \ $ \\
  $11.25\ \ $ & $-2.5000000i$ & $-4.5000000i$ & $\pm0.000001-6.5000000i\ \
  $ & $2.5\ \ $ \\
  $16\ \ $ & $-3.1231056i$ & $-5.1231056i$ &
  $\pm4\times10^{-8}-7.1231056i\ \ $ & $3.1231056\ \ $ \\
  $19.25\ \ $ & $-3.5000000i$ & $-5.5000000i$ &
  $\pm4\times10^{-8}-7.5000000i\ \ $ & $3.5\ \ $ \\
  \hline
\end{tabular}
\end{center}
\vskip 1mm \begin{quote}\small{Table 2. Quasinormal frequencies of
tensor perturbation obtained through numerical calculation when
$m^2\geq3\lambda$.}
\end{quote} \vskip 5mm
\normalsize

\section{Conclusions and Discussions}

We have studied the stability of the BTZ black string against
fermonic perturbation and gravitational perturbation. It is
interesting to see that the BTZ black string can be dynamically
stable provided that the $m^2$, which is determined by the
compactification of the extra dimension, is over a threshold
value, namely $\lambda, 2\lambda, 3\lambda$ respectively for
fermonic perturbation, scalar part of the gravitational
perturbation and the tensor perturbation. When $m^2$ is smaller
than this threshold value, the perturbation can have a growing
mode, which indicates that the BTZ black string can be unstable
and pinch-off to form a black hole.

The situation that the stability and instability coexist
in the dynamical system for the BTZ black string can be understood
from thermodynamical arguments, if we compare the black string
phase with the localized black hole phase with
$\bar{m}A\rightarrow 0$. The entropy argument provided in \cite{7}
shows that the BTZ black string is not a privileged stable phase.
Its entropy equals to that of the localized BTZ black hole with
$\bar{m}A\rightarrow 0$. Our dynamical study supports this
result.

It is of great interest to extend our discussion to the solution
with localized black holes on the brane \cite{7} where the
geometry induced on the two-brane is described by $ds^2=-(\lambda
r^2-k-2\bar{m}A/r)dt^2+(\lambda r^2-k-2\bar{m}A/r)^{-1}dr^2+r^2
d\varphi^2$. The extra term $2\bar{m}A/r$ came from the
four-dimensional nature of the black hole. From the entropy
argument it was argued that when $\bar{m}A$ is deviated from zero,
entropies of black string and black hole will be nonzero and the
black string entropy will exceed that of the black hole for small
$\bar{m}A$. But for large $\bar{m}A$, the black string entropy
will be lower. Thus the black string will first be stable and then
become unstable and pinch-off to form a black hole \cite{7}. The
dynamical description of this process needs careful investigation
and we will report its result elsewhere.

\begin{acknowledgments}

This work was partially supported by NNSF of China, Shanghai
Education Commission and Shanghai Science and Technology
Commission. B.W. would like to acknowledge helpful discussions
with E. Abdalla and the associate programme in ICTP where his work
was completed.
\end{acknowledgments}

\end{document}